\documentclass[prl,aps,twocolumn,showpacs]{revtex4}

\usepackage{graphics}
\usepackage{graphicx}
\usepackage{amssymb,amsmath}
\usepackage{bm}

\newcommand{\mytitle}[1]{

 \twocolumn[\hsize\textwidth\columnwidth\hsize

 \csname@twocolumnfalse\endcsname #1 \vspace{1mm}]}

\newcommand{\beq}{\begin{equation}}
\newcommand{\eeq}{\end{equation}}
\newcommand{\bea}{\begin{eqnarray}}
\newcommand{\eea}{\end{eqnarray}}

\newcommand{\w}{\omega}
\newcommand{\TK}{T_{\rm K}}
\newcommand{\TA}{T_{\rm 1CK}}
\newcommand{\TB}{T_{\rm 2CK}}

\begin{document}

\title{Tunable Kondo-Luttinger systems far from equilibrium}
\author{ C.-H. Chung$^{1,2}$, K.V.P. Latha$^{2}$, K. Le Hur$^{1}$, M. Vojta$^3$, and P. W\"{o}lfle$^{4}$}
\affiliation{
$^{1}$Department of Physics and Applied Physics, Yale University, New Haven, CT 06520, USA \\
$^{2}$Electrophysics Department, National Chiao-Tung University,
HsinChu, Taiwan, R.O.C. \\
$^{3}$Institut f\"ur Theoretische Physik, Universit\"at zu K\"oln, 50937 K\"oln, Germany\\
$^{4}$Institut f\"ur Theorie der Kondensierten Materie, Karlsruher Institut f\"ur Technologie, 76128 Karlsruhe, Germany}
\date{\today}

\begin{abstract}
We theoretically investigate the non-equilibrium current through a quantum dot coupled to
one-dimensional electron leads, utilizing a controlled frequency-dependent
renormalization group (RG) approach. We compute the non-equilibrium conductance for large
bias voltages and address the interplay between decoherence, Kondo entanglement and
Luttinger physics. The combined effect of large bias voltage and strong interactions in
the leads, known to stabilize two-channel Kondo physics, leads to non-trivial
modifications of the conductance. 
For weak interactions, we build an analogy to a dot 
coupled to helical edge states of two-dimensional topological insulators.
\end{abstract}

\maketitle

\begin{figure}[t]
\begin{center}
\includegraphics[width=8.2cm]{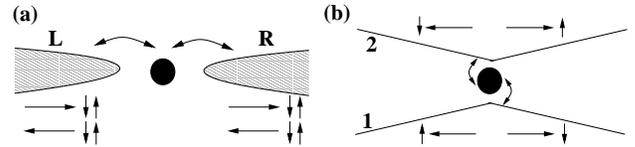}
\end{center}
\par
\vspace{-0.3cm}
\caption{Setups considered in this Letter. (a)
A quantum dot coupled to two 1d Luttinger leads $(L, R)$ where both
left and right-moving electrons in the leads can be either in the spin
up or spin down state. (b) A quantum dot coupled to (helical) edge states
$(1, 2)$ of topological insulators. The right and left moving
electrons on the edges are spin polarized.}
\label{setup}
\end{figure}

Understanding strongly correlated quantum systems far from equilibrium is an
outstanding challenge in condensed-matter physics. Many of the theoretical approaches
that have been proven successful in treating strong correlations are inadequate once the
system is driven out of equilibrium. Quantum dot devices provide an ideal setting to
study transport under non-equilibrium conditions, as they constitute comparatively simple
model systems with high tunability
\cite{Goldhaber,Rosch,Kehrein,Andrei,Boulat,Anders,Mora}. Kondo physics plays a crucial
role in understanding their transport properties \cite{Glazman,Lee}. It has been shown
that several effects in these devices will suppress or modify the Kondo screening, such
as dissipation and the electron-electron interaction in Luttinger liquid quantum wires
that couple to the dot \cite{lehur1,lehur2,Florens,Matveev,Gogolin,Kim}. In this Letter,
we study non-equilibrium currents across quantum dots in nano-settings (see
Fig.~\ref{setup}) involving Kondo entanglement and Luttinger physics \cite{Amir}.

A quantum dot in the Kondo regime coupled to one-dimensional (1d)
leads exhibits either a one-channel Kondo (1CK) or a two-channel Kondo (2CK) ground state
\cite{Gogolin,Kim}, as the Luttinger parameter $K$ is decreased; the control parameter
corresponds to the interaction strength in the 1d leads \cite{Amir}. The non-equilibrium
properties of this system were addressed only in an exactly solvable limit \cite{Lee2} or
in the linear (low bias) region \cite{Ng}. The full crossover in the non-equilibrium
conductance between the 1CK and 2CK fixed points \cite{Gogolin}, with much relevance to
experiments, has not yet been addressed. In particular, interactions in 1d wires are
expected to result in a peculiar non-equilibrium transport \cite{Mirlin,Mason}.

Here, we apply a non-equilibrium RG method \cite{Rosch,chung} to tackle these issues. We
calculate the conductance for bias voltages large compared to the relevant Kondo scales.
We identify signatures of intermediate 2CK behavior in the RG flow for all $K<1$ which
strongly modify the conductance profile. The low-temperature conductance is
non-universal in the sense that it does not depend on $V/\TB$ only, where $\TB$ is the
relevant 2CK scale.

There is also a growing interest in Kondo physics in topological insulator (TI) systems
\cite{Wu,SCZhang,PLee}. Due to spin-orbit coupling,  TIs have gapless helical edge states
where the direction of the electron's spin and momentum are entangled. We shall extend
our analysis to a quantum dot coupled to two helical edges of 2d TIs (see Fig.
\ref{setup}(b)) where it has been shown that 2CK physics is stable even for weak repulsive
electron-electron interactions \cite{PLee}.


{\it Equilibrium properties.}
Let us focus on the setup of Fig. 1(a). We denote by $g_{LR}$ and $g_{LL}=g_{RR}$ the
dimensionless inter-lead and intra-lead Kondo couplings, respectively \cite{Glazman,Lee}.
The RG analysis results in two infrared fixed points \cite{Gogolin,Kim}: the 1CK and 2CK fixed
points. In the former case, all Kondo couplings, $g_{LR}$ and $g_{LL},$ are relevant
under RG transformation and flows towards strong coupling, such that the two leads can be
combined into a single effective lead. In contrast, the 2CK fixed point is reached when
$g_{LR}$ remains small under RG, while $g_{LL}$ grows (and flow to intermediate
coupling). Here, the two leads provide independent screening channels. This 2CK fixed
point is infrared stable for $K<1/2$ (assuming that $g_{LL}=g_{RR}$) \cite{Gogolin,Kim}.

For $K=1$ (free electron leads), where 1CK physics is realized \cite{Glazman,Lee}, the
conductance reaches the unitary limit at low temperatures, $G(T)= 2 G_0
[1-\mathcal{O}((T/\TK)^{2})]$ where $G_0 = e^{2}/h$ is the conductance quantum. Here,
$\TK \equiv D_{0}e^{-1/(g_{LR}^0+g_{LL}^0)}$ is the Kondo temperature; $D_0$ is an
ultraviolet cutoff $(\sim \hbox{few Kelvins})$ whereas $g_{LR}^0$ and $g_{LL}^0$ are the
bare values of the coupling constants. For $T\gg \TK$, from $G(T)\propto g_{LR}(T)^2$,
one finds $G(T) \sim 1/\ln^2 (T/\TK)$.

For all $K<1$, $g_{LR}$ grows slower under RG than $g_{LL(RR)}$.
For $K\ll 1$, one can solve the RG equations
analytically for large $T$. We may neglect $g_{LR}$ in the  RG equation for $g_{LL/RR}$
to obtain the approximate solution $g_{LL/RR}(T) \approx  1/\ln(T/T_K^*)$ with the
shorthand $T_K^* \equiv D_0 e^{-1/g_{LL}^0}$. The coupling $g_{LR}(T)$ is found by
substituting the approximate solution for $g_{LL/RR}(T)$ in the RG equation for the
coupling $g_{LR}$. We evaluate:
\begin{equation}
g_{LR}(T) \approx \frac{(T/D^*)^{\frac{1}{2}(\frac{1}{K}-1)}}{\ln^2(T/\TK^*)}.
\label{gLR-eq}
\end{equation}
where $D^* = \frac{D_0}{[g_{LR}^0
\ln^2(D_0/\TK^*)]^s}$ with $1/s= (1/K-1)/2$. 
We deduce that, for $T\gg T_K^*$, the conductance
$G(T)\propto g_{LR}^2(T)$ essentially follows $T^{1/K-1}/\ln^4(T/T_K^*)$.
Here, the power-law behavior is reminiscent of Luttinger physics whereas the
logarithmic contribution is typical of Kondo correlations.
Importantly, the conductance is {\em not} a universal function
of $(T/T_K^*)$ because transport arises from the sub-leading
coupling $g_{LR}$.

For $1/2<K<1$, the low-temperature physics
is governed by two scales, $\TA < \TB$, with 1CK behavior for
$T \ll \TA$ and 2CK behavior for $\TA \ll T \ll \TB$.
In the limits $K\to1$ and $K\to 0$, we have $\TA=\TK$ and $\TB=\TK^*$,
respectively, with $\TK$, $\TK^*$ defined above.
In general, $\TA \ll \TK$ due to interactions in the leads;
also, $\TB > \TK^*$.
In the presence of particle-hole symmetry, the conductance for
$1/2<K<1$ reaches the
unitary limit as $T\rightarrow 0$. However, potential scattering is
a relevant perturbation with a scaling dimension $(1+K)/2 <1$ 
and causes the conductance to decrease as $T^{1/K-1}$ as $T\rightarrow 0$;
the leading irrelevant operator corresponds to the hopping 
($g_{LR}$) term between the two leads with scaling dimension $(1/K+1)/2$.
Similarly, near the 2CK fixed point reached for $K<1/2$ and $g_{LL}=g_{RR}$,
the leading irrelevant operator ($g_{LR}$ term) 
has dimension $1/2K$ \cite{Kim}, and therefore one
expects $G(T)\propto T^{1/K-2}$ as $T\rightarrow 0$.


\begin{figure}[t]
\begin{center}
\includegraphics[width=8.0cm,height=4.2cm,clip]{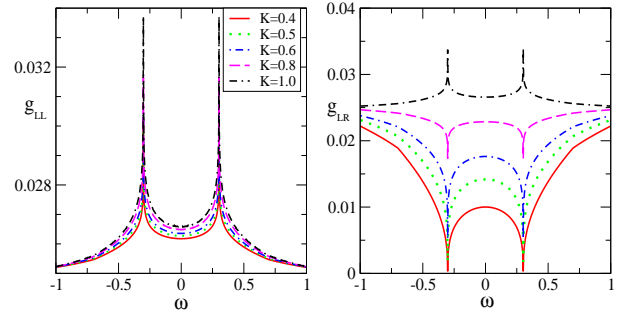}
\end{center}
\par
 \vskip -0.3cm
\caption{
(Color online)
$g_{LL}(\protect\omega)$ and $g_{LR}(\protect\omega)$ for various Luttinger parameters $K$.
The bare couplings are
$g_{\protect\alpha \protect\alpha ^{\prime }}^0=g_{\protect\alpha \protect\alpha}^0=0.025$,
resulting in $T_K^*\sim 4.2\times 10^{-18}$ and $T_K \sim 2\times 10^{-9}$
(in units of $D_0=1$). The bias voltage is $V = 0.6 \gg T_K,T_K^*$.
}
\label{gpergzfigK}
\end{figure}

{\it Non-equilibrium properties.}
We study the low-temperature conductance in the high-bias regime $V\gg \TK,\TK^*$ where
the non-equilibrium RG method can be applied \cite{Rosch}. (In the opposite limit, $V\ll
\TK^*$, we expect the equilibrium results quoted above to be valid after replacing
$T\rightarrow V$.) Setting $\hbar=k_B=e=1$, the non-equilibrium RG equations take the
form:
\begin{eqnarray}
\frac{\partial g_{LL}(\omega )}{\partial \ln D} &=&-\sum_{\beta =-1,1} \left[
g_{L\beta }\left( \frac{\beta V}{2}\right) \right] ^{2}\Theta _{\omega +
\frac{\beta V}{2}}  \nonumber
\end{eqnarray}
\begin{eqnarray}
\frac{\partial g_{LR}(\omega )}{\partial \ln D} &=&-\sum_{\beta =-1,1}\frac{
1}{4}\left[1-\frac{1}{K}\right]g_{LR}\left( \frac{\beta V}{2}\right)\Theta _{\omega +\frac{\beta V}{2}}  \nonumber \\
&-&g_{L\beta }\left( \frac{\beta V}{2}\right) g_{\beta R}\left( \frac{\beta V
}{2}\right)\Theta _{\omega +\frac{\beta V}{2}},
\label{RG}
\end{eqnarray}
where $\Theta _{\omega }=\Theta (D-|\omega +\mathit{i}\Gamma |)$ and  $\beta =-1(+1)$ labels leads $L(R)$.
Further, $\Gamma $ is the decoherence (dephasing) rate at finite bias which cuts off the RG flow
\cite{Rosch}:
\begin{equation}
\Gamma =\pi \sum_{\alpha \alpha ^{\prime }}\int {d\omega f_{\omega }^{\alpha
}\left( 1-f_{\omega }^{\alpha ^{\prime }}\right) [g_{\alpha \alpha^{\prime
}}(\omega )]^{2}},  \label{gamma}
\end{equation}
where the Fermi function obeys $f^{\alpha }(\w)=1/(1+e^{(\w -\mu _{\alpha })/T})$.
We note that there exists an additional contribution
to $\Gamma$ from electron dephasing caused by a finite potential drop in
the Luttinger liquid leads \cite{Mirlin}, which will affect the subleading terms in
$\Gamma$ (given by $g_{LR}(V/2)$).
However, in the low-conductance regime of interest, this voltage drop is small and will
be neglected henceforth.
In general, the perturbative RG approach is valid for $V\gg \TA,\TB$.
In the limit of $V\rightarrow 0$, Eqs. \eqref{RG} reduce to the equilibrium RG equations
(with the flow cut off by temperature), and we recover Eq. \eqref{gLR-eq}.

The renormalized couplings are obtained by self-consistently solving
Eqs.~\eqref{RG} and \eqref{gamma} \cite{Rosch}. As shown in
Fig. \ref{gpergzfigK},
$g_{LL(RR)}(\omega)$ exhibit peaks for all values of $K\leq 1$,
indicating that they grow under RG.
For a given bias voltage, the Kondo coupling $g_{LR}(\omega)$ shows a
crossover from peak to dip structure as $K$ decreases, traducing the
fact that for a fixed bias voltage, $g_{LR}(\omega)$ is either enhanced
or decreased compared to its bare
value $g_{LR}^0$. Let us emphasize that
for sufficiently large bias voltages, as soon as
$K<1$ the coupling $g_{LR}(\omega)$ exhibits a dip close
to $\omega=\pm V/2$, signalling
2CK behavior. The singular behavior at the peaks or dips is cut off by
the decoherence rate (see Eq. (6)), while outside that regime the voltage serves to cut
off the RG flow.

\begin{figure}[t]
\begin{center}
\includegraphics[width=8cm,clip]{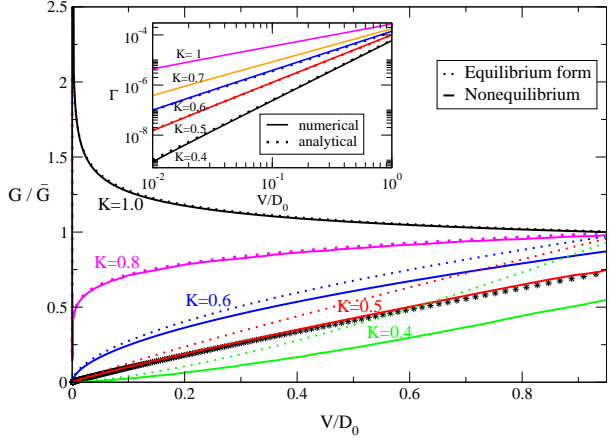}
\end{center}
\vspace{-0.3cm}
\caption{
(Color online)
$G(V)$ normalized to $\bar{G}= 3\pi(g_{LR}^0)^2 / 4$
for various $K$ and Kondo couplings as in Fig. 2.
Here, ``Equilibrium form'' refers to a fit with the expression
$G(V)= 3\pi g_{LR}(\omega=0)^2  /4$ which
consists in substituting $g_{LR}(\omega)$ by $g_{LR}(\omega=0)$
for $|\omega|<V/2$.
For $K=0.5$ we also show the analytical result from Eqs. (7,8) (stars).
Inset: $\Gamma(\omega=0,V)$ for various $K$.
The dashed lines are obtained from the analytical expression in Eq. (9). 
Here, $D_0=1$.
}
\label{Gvlinear}
\end{figure}

From the Keldysh calculation up to second order in the tunneling amplitudes,
the current reads:
\begin{eqnarray}
I =\frac{3\pi}{4}\int d\omega \Big[\sum_{\sigma }g_{LR}(\omega
)^{2} f_{\omega }^{L}(1-f_{\omega }^{R})\Big]-(L\leftrightarrow R).
\label{current}
\end{eqnarray}
For small bare couplings $g_{\protect\alpha \protect\alpha ^{\prime
}}^0=g_{\protect\alpha \protect\alpha }^0$ this
perturbative calculation of $I$ remains valid for $V> \TK^*$, 
implying at high bias voltage $V\gg T$ contributions to the current 
over a frequency window $-V/2< \omega < V/2$.
For bias voltages $\TK^*< V \ll D_0$, with decreasing $V$
we find the differential conductance $G(V)\equiv dI/dV$ 
approaches the equilibrium form of the
conductance $G(T\rightarrow V) \propto
V^{1/K-1}/\ln^4(V/\TK^*)$ with $G(T)\propto g_{LR}(T)^2$
(see Eq. (\ref{gLR-eq}) and Fig. \ref{Gvlinear}).

In the remainder, we analyze $G(V)$ for larger bias voltages.
For $K=1$, we checked that the nonlinear conductance satisfies
$G(V)\propto 1/\ln
^{2}(V/\TK)$ for $V\gg\TK$ \cite{Kaminski}.
Here, one can replace $g_{LR}(\omega)$
by $g_{LR}(\omega=0)\approx g_{LR}(T\rightarrow V)$.
When decreasing $K$, the double peak
structure in $g_{LR}(\omega)$ at $\omega=\pm V/2$
turns progressively into dips which
acquire a complex shape as a result of the decoherence rate
$\Gamma$ and the
electron-electron interaction which hinders the inter-lead
electron tunneling. The effect
becomes more pronounced for small $K$ values
associated with the 2CK fixed
point, rendering the ``flat'' approximation
$g_{LR}(\omega)\approx g_{LR}(\omega=0)$ not
justified; see Fig.  \ref{Gvlinear}.

To gain an analytical understanding of the
small-$K$ non-equilibrium regime, we may treat
$g_{RL}(\omega)$ within the interval $-V/2<\omega<V/2$
as a semi-ellipse \cite{chung}.
The current $I$ reads
\begin{equation}
I \approx \frac{3\pi}{4} \left[\frac{\pi}{4}g_{LR}(\omega=0)^2 +
\left(1-\frac{\pi}{4}\right)g_{LR}(\omega=V/2)^2\right].
\end{equation}
For $K< 1$, we manage to obtain an approximate analytical
form for the couplings $g_{LR}(\omega=V/2)$ and $g_{LR}(\omega=0)$.
Solving Eqs.~\eqref{RG} in the limit $D\rightarrow 0$, we find:
\begin{eqnarray}
g_{LR}(\omega=0) &\approx& g_{LR}(T\rightarrow V) {\cal F}(K) \\ \nonumber
g_{LR}\left(\frac{V}{2}\right) &\approx& \frac{4  \left(\frac{\Gamma V}{D^{*2}}\right)^{\frac{1}{4}(\frac{1}{K}-1)}}{\ln^2\left(\frac{\Gamma V}{(\TK^*)^2}\right)},
\end{eqnarray}
where $g_{LR}(T\rightarrow V)$ is the equilibrium form of $g_{LR}$
in Eq. \eqref{gLR-eq} with $T$ replaced by $V$,
and we have defined ${\cal F}(K)= 2^{1+ \frac{1}{4}(1-1/K)}-1$, with ${\cal F}(K=1)=1$.
Using Eqs. (5) and (6), we obtain a closed expression for the conductance:
\begin{eqnarray}
G(V) &\approx& \frac{3\pi^2}{16} \left(\frac{V}{D^*}\right)^{\frac{1}{K}-1}{\cal R}(V) \\ \nonumber
&+& 12\pi  \left(1-\frac{\pi}{4}\right) {\cal W}'(V),
\end{eqnarray}
where ${\cal W}'(V) = d{\cal W}/dV$ and:
\begin{eqnarray}
{\cal R}(V) &=& {\cal F}(K)^2 \left[\frac{1/K}{\ln^4\left(\frac{V}{\TK^*}\right)} - \frac{4}{\ln^5\left(\frac{V}{\TK^*}\right)}\right] \\ \nonumber
{\cal W}(V) &=& \frac{V(\Gamma V/D^{*2})^{\frac{1}{2}(\frac{1}{K}-1)}}{\ln^4[\Gamma V/(\TK^*)^2]}.
\end{eqnarray}
For completeness, we have kept the less dominant contribution $\sim 1/\ln^5(V/\TK^*)$ in
${\cal R}(V)$. To rigorously define the function ${\cal W}(V)$, we need to provide an
analytical expression for the decoherence rate in Eq. (3).
\begin{figure}[t]
\begin{center}
\includegraphics[width=7.7cm,clip]{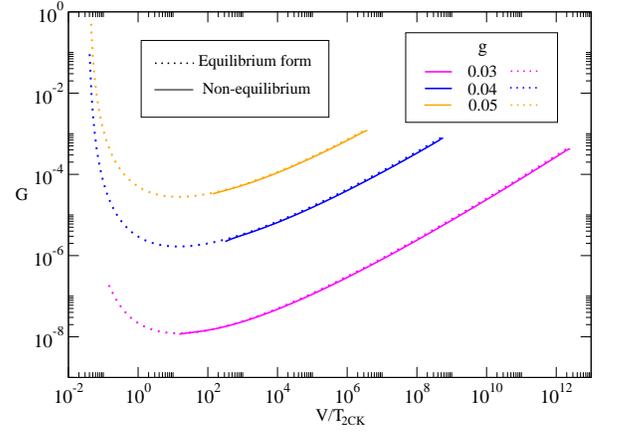}
\end{center}
\vspace{-0.3cm}
\caption{
(Color online)
$G$ versus $V/\TB$ for $K=0.6$ for various initial Kondo
couplings $g_{\protect\alpha \protect\alpha ^{\prime }}^0=g_{\protect\alpha
\protect\alpha}^0=g$ (in units of $D_0=1$).
}
\label{GVscaling}
\end{figure}
Using an analogous reasoning as for the non-equilibrium current $I$, to second order in $g_{LR}^0$, we extract
\begin{equation}
\Gamma \approx  \frac{\pi^2}{4} {\cal F}(K)^2 \frac{V\left(\frac{V}{D^*}\right)^{\frac{1}{K}-1}}{\ln^4\left(\frac{V}{T_K^*}\right)}.
\end{equation}
Close to $K=1/2$, we can safely neglect contributions in $(g_{LR}^0)^{2+(1/K-1)}$ and therefore to second order in $g_{LR}^0$, we find $\Gamma \approx
(\pi^2/4)[g_{LR}(\omega=0)]^2$.
We have checked our analytical expression of $\Gamma$
against a numerical treatment of Eqs. (2) and (3); see inset in Fig. \ref{Gvlinear}.
Notably, the decoherence rate contributes to a ``distinct'' power law
$\sim V^{\frac{1}{2}\left(\frac{1}{K^2}-1\right)}$ in
the non-equilibrium conductance $G(V)$ where $(1/K^2-1)/2>1/K-1$ for $K<1$,
rendering the second term in Eq. (7) to be
subleading. The conductance becomes smaller than
its equilibrium counterpart since
$g_{LR}(\omega=V/2)< g_{LR}(\omega=0)$.
A comparison between the analytical formula in
Eq. (7) and the numerical integration of Eqs. (2-4) is shown in
Fig. \ref{Gvlinear}. As our results are based on one-loop RG, we may expect both corrections
to the power-law prefactors and further
subleading terms upon including higher-loop contributions.

Our results show that $G(V)$ for voltages $\TB\ll V \ll D_0$
is {\em not} an universal function of $V/\TB$ (even for fixed $K$):
Fig.~\ref{GVscaling} displays $G$ versus $V/\TB$
for various initial Kondo couplings, with $\TB$ extracted from the RG flow.
As it becomes also clear from Eqs.~(7,8), the non-equilibrium conductance for $V\gg\TB$
is a function of both $V/D_0$ and $V/\TK^*$, and hence has a non-universal profile.
This is again related to the fact that transport arises from the subleading
coupling $g_{LR}$.


{\it Topological insulators.}
We can extend these results to a quantum dot coupled to helical edges of 2d TIs; see Fig. 1(b).
In contrast with the setup of Fig. 1(a) where
single-particle backscattering terms
caused by the quantum dot cut the system into two separate parts,
such backscattering terms are now forbidden due to time-reversal
symmetry of the helical edges \cite{Wu,PLee}.
Following Refs. \onlinecite{Hou,Teo}, the two edges of spinless electrons
can be mapped onto spinful Luttinger liquids with distinct Luttinger parameters for charge
($K_c$) and spin ($K_s$) degrees of freedom: $K_c= \bar{K}$, and
$K_s=1/\bar{K}$ with $\bar{K}$ being Luttinger parameter for the
helical edges.
In equilibrium, the system
is equivalent to an anisotropic 2CK model with the Kondo couplings
$g_{\alpha\beta}^{xy,z}$ ($\alpha, \beta=1,2$) \cite{PLee} and in contrast 
to the case in Fig. 1(a), the
2CK fixed point is stable for $\bar{K}<1$.
In the limit of $\bar{K}\to 1^-$ where the anisotropy in the Kondo couplings
is negligible, we have checked that the RG scaling
equations for the Kondo couplings $g_{11/22}$ and $g_{12}$
are identical to those of the Kondo couplings $g_{LL/RR}$ and $g_{LR}$.
We identify:
$g_{LL/RR}\to g_{11/22}$, $g_{LR}\to g_{12}$,
$ \frac{1}{K} \to \bar{K}+1/\bar{K}-1$ \cite{RGTI}. Our results on
non-equilibrium transport across a quantum dot coupled to weakly interacting 
Luttinger leads
are directly applicable.


{\it Summary.}
We have studied non-equilibrium transport through a Kondo dot
coupled to Luttinger-liquid leads and calculated the conductance profile at bias voltages
larger than the Kondo scales of the system.
The RG flow at large bias shows signatures of intermediate 2CK physics
for all Luttinger parameters $K<1$.
As the conductance $G$ arises from the coupling $g_{LR}$ which is subleading,
$G(T\to 0)$ is not a universal function of $V/\TB$ as it also depends 
on $V/D_0$. Our results push forward
the knowledge of correlation effects in nanosystems far from
equilibrium and should
stimulate further experimental works on transport through dots
coupled to quantum wires
and carbon nanotubes. We have also shown that our theoretical
framework is applicable to
the Kondo effect at the helical edges of topological insulators.

We thank A. Rosch for many helpful discussions.
This work is supported by the NSC grant
No.98-2918-I-009-06, No.98-2112-M-009-010-MY3, the MOE-ATU program, the
NCTS of Taiwan, R.O.C. (C.H.C.), the Department of
Energy in USA under the grant DE-FG02-08ER46541
(K.L.H.), and the DFG via SFB 608, SFB/TR-12 (M.V.),
and the Center for Functional Nanostructures (P.W.). KLH thanks the kind hospitality of LPS Orsay.



\begin{thebibliography}{}
\bibitem{RGTI}
The scaling dimensions of the couplings are
$[g_{11/22}^{xy}]=\bar{K}$, $[g_{11/22}^{z}]=1$,
$[g_{12}^{xy}]=[g_{12}^z]\equiv \delta = 1/2(\bar{K}+1/\bar{K})$.
\end{thebibliography}
\end{document}